\documentclass[preprint]{emulateapj} 

\shorttitle{Puffy Magnetic EGP's} 
\shortauthors{Batygin \& Stevenson} 

\begin{document}
 
\title{Inflating Hot Jupiters With Ohmic Dissipation}  
\author{Konstantin Batygin \& David J. Stevenson } 

\affil{Division of Geological and Planetary Sciences, California Institute of Technology, Pasadena, CA 91125} 

\email{kbatygin@gps.caltech.edu}
 
 \begin{abstract}
We present a new, magnetohydrodynamic mechanism for inflation of close-in giant extrasolar planets. The idea behind the mechanism is that current, which is induced through interaction of atmospheric winds and the planetary magnetic field, results in significant Ohmic dissipation of energy in the interior. We develop an analytical model for computation of interior Ohmic dissipation, with a simplified treatment of the atmosphere. We apply our model to HD209458b,  Tres-4b and HD189733b. With conservative assumptions for wind speed and field strength, our model predicts a generated power that appears to be large enough to maintain the transit radii, opening an unexplored avenue towards solving a decade-old puzzle of extrasolar gas giant radius anomalies. 
 \end{abstract}
 
 
 \section{Introduction}

The detection of the first transiting extrasolar planet HD209458b \citep{2000ApJ...529L..45C,2000ApJ...529L..41H} marked the first observation of a planet whose radius is anomalously large. With the current aggregate of transiting planets exceeding 60, over-inflated ``hot Jupiters" are now known to be common (Fig.1), and understanding their radii has become recognized as an outstanding problem in planetary astrophysics \citep{2010RPPh...73a6901B}. Most proposed explanations require an interior power source that would replace the radiated heat from gravitational contraction and cause a planet to reach thermal equilibrium with a larger-than-expected radius.  In the context of such solutions, the generated heat must be deposited into the interior envelope, i.e. below the radiative/convective boundary, in order to maintain the core entropy (and therefore the radius) of the planet. Notably, eccentricity tides \citep{2001ApJ...548..466B}, obliquity tides of a Cassini state \citep{2005ApJ...628L.159W}, and deposition of  kinetic energy to adiabatic depths by dynamical and convective instabilities \citep{2002A&A...385..156G} have been invoked to provide an extra power source in the interior of the planet. It has been shown that the required powers are rather modest \citep{2007ApJ...661..502B}, but it is unlikely that any of the proposed solutions alone are able to account for all observed radii \citep{2010RPPh...73a6901B,2009SSRv..tmp..108F}. 

Here we show that the anomalous sizes of close-in exo-planets can be explained by a magnetohydrodynamic mechanism. The interactions of zonal winds with the expected planetary magnetic field in a thermally ionized atmosphere induce an emf that drives electrical currents into the interior. These  currents dissipate Ohmically and thus maintain the interior entropy of the planet. The primary controlling factors in our model are the atmospheric temperature, wind velocity and strength of the magnetic field, as they dictate how much current is allowed to penetrate the interior. Other variables, such as metalicity also contribute, but to a smaller degree. Our results predict that interior heating of this kind occurs in all close-in exoplanets with magnetic fields, but it is negligible if the atmospheric temperature is not high enough for sufficient thermal ionization to take place. Smaller, but hot exoplanets are attributed to heavy element enrichment in the interior. While the inflation mechanism we present here is general, the quantitative modeling in this work is specific to HD209458b, Tres-4b, and HD189733b which are arguably the better studied transiting exoplanets.

\begin{figure}
\epsscale{1.0}
\plotone{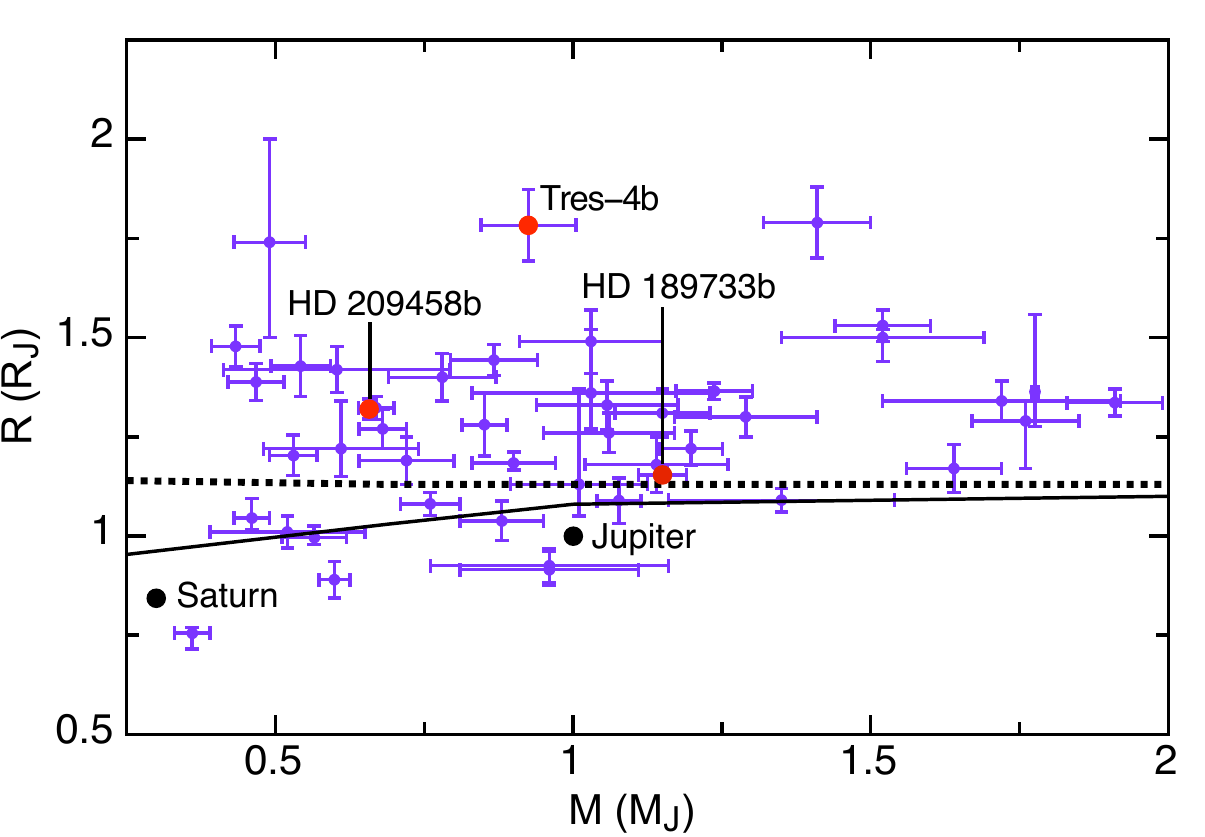}
\caption{Scatter-plot of mass vs. radius of transiting Jovian exo-planets. The three planets considered in the text as well as Jupiter \& Saturn are labeled. The two lines represent the theoretical mass-radius relationships for a core-less planet (dashed) and one with a $40M_{\oplus}$ core (solid) from \cite{2003ApJ...592..555B}. Planets above the dashed line require an inflation mechanism to halt gravitational contraction.}
\end{figure}

\section{Structural Model \& Electrical Conductivity} 

Unlike Jupiter and Saturn, close-in extra-solar gas giants are exposed to high irradiation due to their proximity to parent stars. This forces their atmospheric temperature-pressure profiles to be significantly shallower than their solar system counterparts \citep{2009SSRv..tmp..108F}. In particular, the lower atmospheres ($P\gtrsim0.1$ Bars) of hot Jupiters are believed to be almost isothermal while the radiative/convective boundaries are thought to lie at $P\sim100-1000$ Bars, depending on the planet \citep{2008ApJ...682..559S}. 

The isothermal sections of extrasolar gas giant atmospheres often reach temperatures close to 2000 K \citep{2009ApJ...699.1487S} and in some cases, even higher \citep{2009Sci...325..709B}. These temperatures are not high enough to ionize H or He significantly, however, alkali metals such as Na and K will be partially ionized. As a result, electrical conductivity in the interior of a hot Jupiter is dominated by ionization of hydrogen, while in the outer region of the planet, electrical conductivity is primarily due to the ionization of alkali metals, with the transition between the two inoization regimes taking place at $P\sim300$ Bars. 

Thermal ionization is governed by the Saha equation:
\begin{equation}
\frac{n_{j}^{+}n_{e}}{n_{j}-n_{j}^{+}}=\left(\frac{m_{e}k_{b}T}{2\pi\hbar^2}\right)^{\frac{3}{2}}\exp\left(-I_{j}/k_bT\right),
\end{equation} 
where $n_{j}$ and $n_{j}^{+}$ are the total and positively ionized number densities of constituent $j$ respectively, $n_{e} = \sum n_{j}^{+}$ is the total electron number density, $m_{e}$ is the electron mass, $k_{b}$ is Boltzmann's constant, $T$ is temperature, $\hbar$ is Plank's constant, and $I_{j}$ is the ionization potential of constituent $j$. If the ionization is far from complete ($n_j^{+}\ll f_jn$), the abundances of alkali metals, $f_j$ are held constant, and the atmosphere is isothermal, it is easy to show that the electron number density takes on an exponential profile with an ionization scale-height that is twice as large as the density scale-heght:
\begin{equation}
n_e=n_0\sqrt{\sum_{i=1}^{N}f_i\chi_i}e^{\frac{r_0-r}{2H}},
\end{equation}
where $\chi$'s are the RHS's of equation (1), $r_0$ is the radial distance at some reference point ($P=10$ Bars) and $H=k_bT/\mu g$ is the density scale-height. In our ionization calculations, we considered the following alkali metals: Na, K, Li, Rb, Fe, Cs and Ca. Their abundances and ionization potentials were inferred from \cite{1999ApJ...519..793L} and \cite{2000asqu.book..499C} respectively.

The atmospheric temperatures above the isothermal layer differ significantly from planet to planet. In particular, thermal inversions have been detected in the atmospheres of  HD209458b \citep{2007ApJ...668L.171B} and Tres-4b \citep{2009ApJ...691..866K} but not in HD189733b. In our models, we adopt atmospheric temperature profiles similar to that of \cite{2009ApJ...699.1487S} for HD209458b and Tres-4b, and the 1D profile of \cite{2010ApJ...709.1396F} for HD189733b. The relatively cool temperatures attained above P $\lesssim0.1$ Bars are of significant importance to our models because they provide insulating shells which are impenetrable to radial current. Consequently, current loops are necessarily setup through the interior, and any current flowing in the ionosphere is not relevant. We place the radiative/convective boundary at $ P\sim100$ Bars in all of our models. 

We did not have to explicitly compute the ionization fractions of H and He, as they are published in the equation of state \citep{1995ApJS...99..713S}, which we employed in our model. In particular, we used the "interpolated" version of the equation of state, where ionization occurs smoothly with pressure and temperature. Although the planetary structure was core-less, we mimicked the presence of a core by changing the Helium content from $Y=0.24$ to $Y=0.3$ \citep{2003ApJ...594..545B} in some of our models.

Having computed the electron number density, the electrical conductivity of a gas is given by \citep{2002moph.book.....T}
\begin{equation}
\sigma=\frac{n_{e^{-}}}{n}\frac{e^2}{m_{e}A}\sqrt{\frac{\pi m_e}{8k_bT}}
\end{equation}
where $n$ and $A$ are the number density, and the number density weighted cross-section of everything other than electrons. Strictly speaking, the above equation is only valid for non-degenerate gas. However, by the point matter becomes degenerate in our models, the resistivity is completely negligible and the details of its profile have no noticeable effect on the results.

Since we are only interested in the part of the planet, interior to the atmospheric temperature minimum, we define the model radius $r=R$ as the point of maximal conductivity in the atmosphere ($P = 75$ mbars), and we set the outer edge of our model at the conductivity minimum, $r=R+\gamma$ ($P=30$ mbars). We place the bottom boundary of the ``weather" layer of the atmosphere at a pressure of $P=10$ Bars and denote it as $r=R-\delta$. Consequently, the ``inert" layer of the atmosphere is between $100\lesssim P\lesssim10$ Bars. A computed electrical conductivity profile for HD209458b is presented in figure (2), along with a simplified conductivity profile resulting from equation (2). Because the  functional profiles (dashed curve) are in good agreement with the numerically computed profile, we utilize them in all future calculations (see appendix).

\begin{figure}
\epsscale{1.0}
\plotone{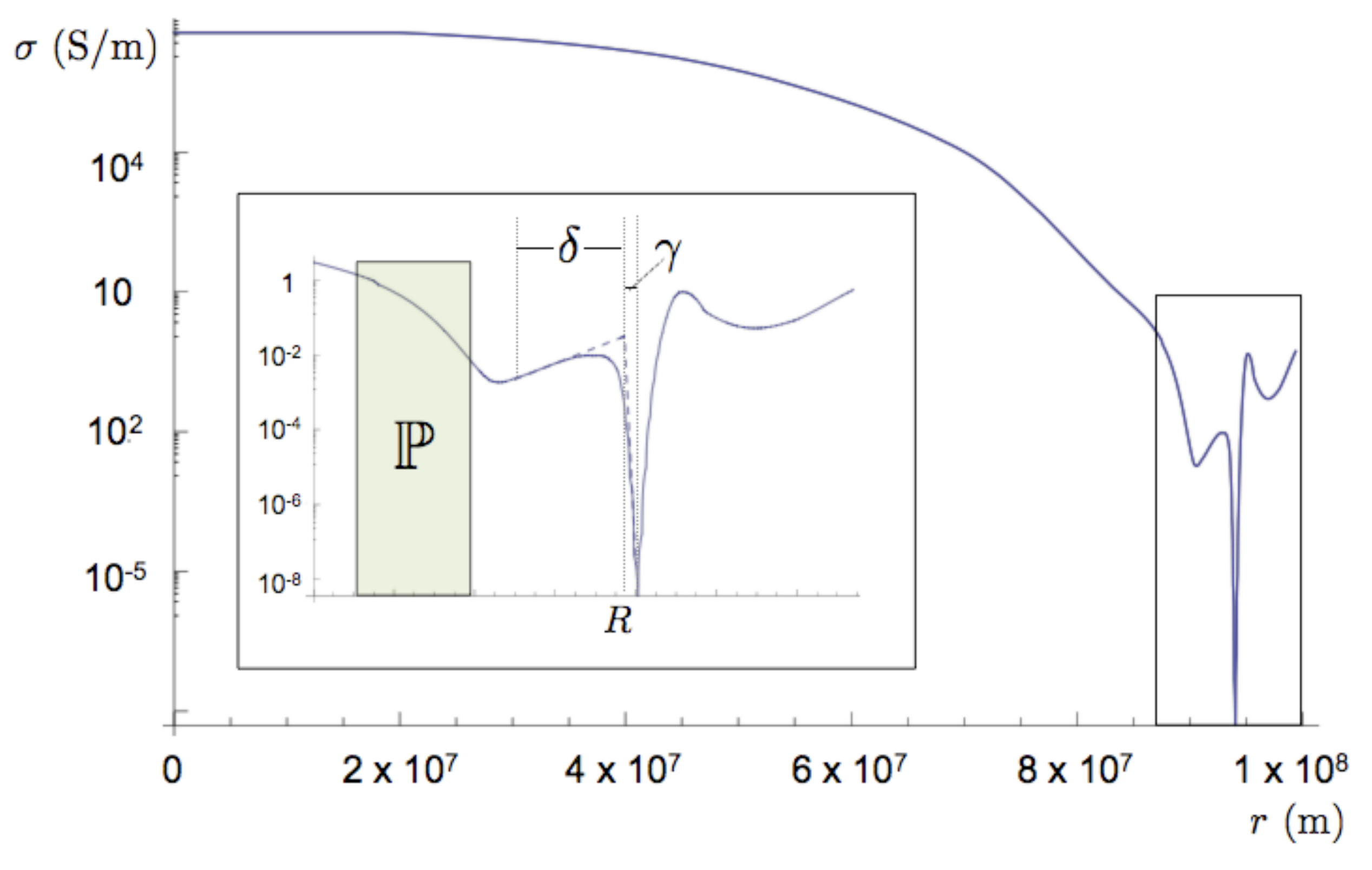}
\caption{Electrical conductivity profile of the nominal HD209458b model with $T_{iso}=1700$K, $Y=0.24$ and $Z=1\times$solar. The inset is a magnification of the profile in the outer part of planet. The model parameters $R,\delta$ and $\gamma$ are labeled. The dashed lines are functional approximations to the conductivity profile where zonal flow is present. The highlighted region corresponds to the upper convective envelope ($100-3000$)Bars, where most of the interior dissipation takes place. }
\end{figure}

\section{Analytical Theory}

Global circulation models \citep{2008ApJ...682..559S,2009ApJ...699..564S,2008ApJ...674.1106L,2009ApJ...700..887M} have shown that winds on hot Jupiters, specifically HD209458b and HD189733b, can attain velocities of order $v\sim1$km/s. It appears that two qualitative wind patterns are present. In the upper atmosphere ($P\lesssim30$mbars), wind flows from the sub-stellar point to the anti-stellar point symmetrically across the terminator. Deeper down, a strong eastward zonal jet develops. Importantly, the development of zonal jets have been observed in virtually all simulations (see \citet{2009arXiv0911.3170S} for a comprehensive review). 

Prior to obtaining a formal solution to the problem, we can identify some of its features. First, if the planet's dipole moment is aligned with the rotation axis and we consider only zonal flow, then there is azimuthal symmetry. Second, it is immediately apparent from the geometry of the zonal jet and the dipole field that the induced current will be meridional. In the atmosphere, we expect the current to flow from the poles to the equator where it penetrates the interior of the planet and completes the loop (Fig 3). 

The general induction equation can be written as:
\begin{equation}
\frac{\partial\vec{B}}{\partial t}=-\vec{\nabla}\times\lambda\left(\vec{\nabla}\times\vec{B}\right)+\vec{\nabla}\times\left(\vec{v}\times\vec{B}\right),
\end{equation}
where  $\vec{B}$ is the magnetic field and $\lambda\equiv 1/\mu_0\sigma$ is the magnetic diffusivity \citep{1978mfge.book.....M}. We express the magnetic field as a dipole background component and an induced component: $\vec{B}=\vec{B}_{dip}+\vec{B}_{ind}$ with $\nabla\times\vec{B}_{dip}=0$. This assumes no dynamo generation in the region. The induced magnetic field will tend to point in the same direction as the velocity field, so we can make the approximation $\vec{v}\times\vec{B}\approx\vec{v}\times\vec{B}_{dip}$. We assume that the prescribed velocity field and background magnetic field are not strongly modified by the induced field i.e. $Rm\equiv vL/\lambda\lesssim1$, an assumption satisfied in our models with T$\leqslant1700$K. Finally, we seek a steady-state solution, so we require $\partial\vec{B}/\partial t=0$. With these assumptions, the induction equation simplifies to:
\begin{equation}
\vec{\nabla}\times\lambda\left(\vec{\nabla}\times\vec{B}_{ind}\right)=\vec{\nabla}\times\left(\vec{v}\times\vec{B}_{dip}\right).
\end{equation}
We can "uncurl" this equation and use Ampere's law $\vec{\nabla}\times\vec{B}=\mu_0\vec{J} $ to recover Ohm's law: 
\begin{equation}
\overrightarrow{J}_{ind}=\sigma\left(\overrightarrow{v}\times\overrightarrow{B}_{dip}-\vec{\nabla}\Phi\right).
\end{equation}
By continuity, $\nabla\cdot\vec{J}$ must vanish. As a result, 
\begin{equation}
\overrightarrow{\nabla}\cdot\sigma\overrightarrow{\nabla}\Phi=\overrightarrow{\nabla}\cdot\sigma\left(\overrightarrow{v}\times\overrightarrow{B}_{dip}\right).
\end{equation}
If the conductivity takes on an exponential form, there exists an analytical solution for $\Phi$ and in our models, we confine the atmospheric flow to the region where conductivity is exponential. In the interior region, the electric potential is also governed by the above equation, with the right-hand side is set to zero. However, since the interior conductivity does not take on a simple analytical form, the above equation there must be solved numerically. 

We take a nominal value for the "strength" of the field at the surface of the planets to be $||B||_R=10^{-3}$T, approximately the value expected from scaling the field via the Elsasser number $\Lambda\equiv\sigma B^2/2\rho\Omega\sim1$, where $\Omega$ is the planetary rotation rate (assumed tidally locked). The magnetic field scaling argument based on energy flux also suggests a similar value \citep{2009Natur.457..167C}. For comparison, Jupiter's surface field is $||B||_{R_{jup}}=4.2\times10^{-4}$T \citep{2003E&PSL.208....1S}. We approximate the zonal wind as $v\propto v_m\sin(\theta)\hat{\phi}$ where $v_m$ is the maximum speed attained by the wind and set $v_m=1$ km/s (see appendix for more details). 

Once we have the solution for the current, we can compute the total Ohmic dissipation rate below some radius $r$:
\begin{equation}
\mathbb{P}=\int\int\int\frac{\vec{J}^2}{\sigma(r)}dV.
\end{equation}
In order to satisfy continuity, the magnitude of the current density must be constant along its path in the interior. As a result, it is apparent from the above equation that most of the dissipation takes place in the upper layers of the planet, where conductivity is not too great, and the solution is insensitive to the details of the conductivity profile in the deep interior, as long as it remains high. The Ohmic heat that is generated in the convective envelope of the planet replaces gravitational contraction, and is lost by radiative cooling at the radiative/convective boundary. Consequently, to ensure a null secular cooling rate, we need the Ohmic dissipation rate to at least compensate for the the radiative heat flux at the radiative/convective boundary \citep{1968psen.book.....C}. 

\begin{figure}
\epsscale{1.0}
\plotone{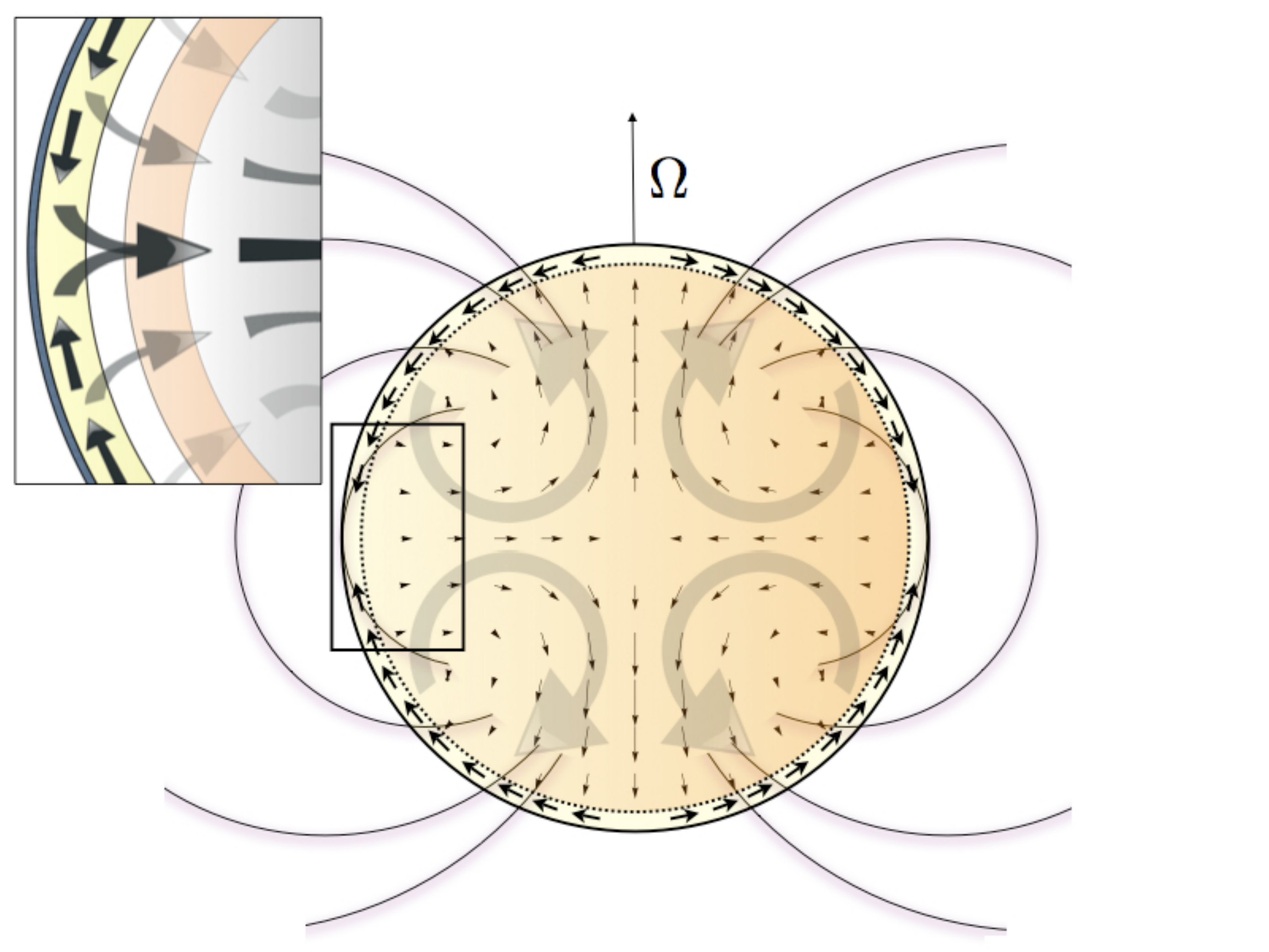}
\caption{Side view cross-section of induced current due to zonal wind flow. The interior vector field, plotted with small arrows, is a quantitative result of the model. The large semi-transparent arrows are illustrations. The yellow shell in the inset represent the region to which we confine the zonal flow (10-0.03 Bars). The orange region denotes the region of interior heating.}
\end{figure}

\section{Model Results}

It has been shown that extra-solar gas giants require between $10^{-6}$ and $10^{-2}$ of the irradiation they receive to be deposited into the adiabatic interior to maintain their radii \citep{2001ApJ...548..466B,2007ApJ...661..502B, 2009arXiv0910.5928I}, although the exact number depends on the metallicity of the atmosphere and the mass of the heavy element core in the interior of the planet. Under the assumption of solar metallicity and no core, HD209458b requires $3.9\times10^{18}$W, Tres-4b requires $8.06\times10^{20}$W and HD189733b requires no heating at all \citep{2007ApJ...661..502B,2009arXiv0910.5928I}. Within the context of our model, HD209458b and HD189733b are easily explained. To adequately explain Tres-4b however, we require an enhanced (10$\times$solar) metallicity in the atmosphere to reduce the required heating down to $5.37\times10^{19}$W.

Table (1) presents a series of models with various temperatures, Helium contents, and metallicities of the planets under consideration. Upon inspection, it is apparent that the global heating rate scales exponentially with temperature, and as a square root of the metallicity. Both of these scalings can be easily understood by noting that scaling the conductivity profile by a multiplicative factor causes a corresponding change in dissipation while equations (1) and (2) relate temperature and metallicity (i.e. $f$) to the conductivity.

It is also noteworthy that the models with a simulated core produce approximately the same amount of heating as the coreless models. This is because most of the dissipation takes place in a region where ionization of alkali metals still dominates the electrical conductivity and the somewhat hotter interior isentrope makes little difference - recall that the value of the conductivity is unimportant in the deep interior.

There are a number of other scalings present within our model. For instance, the total Ohmic dissipation rate is proportional to the squares of the wind speed and the strength of the magnetic field, $\mathbb{P}\propto(B/10^3$T$)^2(v_m/1$km s$^{-1})^2$. Additionally, to leading order, the dissipation in the atmosphere scales linearly with the thickness of the atmosphere, while the interior dissipation approximately scales quadratically. Consequently, along with the conductivity effect, hotter atmospheres also lead to more dissipation by virtue of a physically larger atmosphere.

It is important to understand that Ohmic heating does not only affect the interior. Because the induced current's ``return path" lies in the atmosphere (Fig. 3), the atmosphere also gets heated. This heating, along with magnetic drag on the flow are the limiting factors of our theory. 

Consider the nominal case of HD209458b with $T_{iso}=1700$K and $Z=1\times$solar. In this model, the heating required to inflate the planet is attained at a depth of $\sim90$ Bars, essentially \textit{at} the radiative/convective boundary. Ohmic heating in the atmosphere is small (only $\sim4\%$) in comparison with the irradiation, having little effect on the planet's evolution or structure and $R_m\sim0.3$. In other words, the assumptions implicit in our calculation are satisfied and the mechanism seems to explain the transit radius adequately. However, if we go to the model with $T_{iso}=2000$K, $R_m\sim3$, the Ohmic dissipation in the atmosphere is comparable with the insolation, and the assumptions of our model may no longer apply. 

The nominal model of Tres-4b with $T_{iso}=2250$K also runs into the same problem. Here, $R_m\sim15$, and the Ohmic dissipation in the atmosphere is again comparable with the insolation. However, if we imagine that magnetic drag reduces the wind velocity by a factor of $\sim3$, our results fall in the right ballpark to explain Tres-4b's radius, in the scenario where its atmospheric opacity is super-solar. Finally, consider the model of HD189733b. For this configuration of parameters, our mechanism does not predict a significant amount of Ohmic dissipation at adiabatic depths, consistent with an un-inflated radius. A similar scenario is observed for the model of HD209458b with $T_{iso}=1400$K. Overall, it appears that within the current setup of the model, the cumulative heating below the weather layer of the atmosphere i.e. $r<R-\delta$ is of order a few$\times10^{-2}$ of the heating that takes place in the atmosphere. Provided that this ratio of magnitudes holds up in a more dynamical treatment of the problem, it can provide an upper limit to the maximum inflation that can be explained with Ohmic dissipation.

\section{Discussion}
In this letter, we have presented a new, magnetohydrodynamic mechanism for inflation of extrasolar gas giants. Our calculations show that the heating, necessary to maintain the seemingly anomalous radii of transiting exo-planets, naturally emerges from considerations of interactions between partially ionized winds and the planetary magnetic field. Interestingly, there seems to be a set limit to the extent that Ohmic dissipation can heat the interior, making this theory testable. Currently, there is significant uncertainty with respect to the calculation of the required interior heating, because core masses are unknown. However, dynamical determinations of interior structure \citep{2009ApJ...704L..49B,2009ApJ...698.1778R} may allow us to resolve the degeneracy for a fraction of observed planets, and provide a solid test-bed for the mechanism we've presented here.

There is a number of interesting additional questions that our model inevitably brings up. First, recall that our treatment of the induction equation is kinematic. In reality, flow modification by the Lorenz force may play an important role in determining the actual wind patterns. While this effect may be small for HD209458b and HD189733b, weather on hotter planets, such as Tres-4b or Wasp-12b may be more intimately linked with their magnetic fields, calling for a magentohydrodynamic treatment of the atmospheric circulation. Generally, when zonal winds interact with a background dipole field, they give rise to poloidal current which in turn, gives rise to a predominantly toroidal, unobservable field. However, the dayside-to-nightside flows that are present at higher levels in the atmosphere may modify the flow in an interesting way that may eventually be astronomically observable. 

Second, we are neglecting the stellar magnetic field. The star's magnetic field is likely to be considerably smaller than the planetary field at the planetary orbital radius, but induction by stellar field as well as coupling of the stellar and planetary magnetic field lines is certainly plausible. This too, may produce an astronomically observable signature. Finally, we are neglecting the effects the induced current in the interior on the planetary dynamo. Considerations of this sort may influence the background magnetic field of the planet. All of these aspects call for a self-consistent treatment of the full problem. Such calculations would no-doubt provide further insight into the physical structure of extra-solar gas giants. 

\begin{appendix}

\begin{table}
\begin{center}
\caption{Ohmic dissipation acquired at various pressures in various models of HD209458b, Tres-4b, and HD189733b}
\begin{tabular}{llllllll}
\tableline\tableline
Planet & Y & $T_{iso}$ (K) & Z ($\times$solar) & $\mathbb{P}$ [$P<10$ Bars] (W) & $\mathbb{P}$ [$P>10$ Bars] (W) & $\mathbb{P}$ [$P>100$ Bars] (W) \\
\tableline
HD209458b & 0.24 & 1400 & 1   & $2.30\times10^{19}$ & $2.23\times10^{17}$ & $1.09\times10^{16}$  \\
HD209458b & 0.24 & 1400 & 10 & $7.28\times10^{19}$ & $7.06\times10^{17}$ & $3.43\times10^{16}$  \\
HD209458b & 0.24 & 1700 & 1   & $1.14\times10^{21}$ & $1.01\times10^{19}$ & $5.60\times10^{17}$  \\
HD209458b & 0.24 & 1700 & 10 & $3.61\times10^{21}$ & $3.19\times10^{19}$ & $1.77\times10^{18}$  \\
HD209458b & 0.24 & 2000 & 1   & $1.22\times10^{22}$ & $3.24\times10^{20}$ & $7.09\times10^{19}$  \\
HD209458b & 0.24 & 2000 & 10 & $3.89\times10^{22}$ & $1.05\times10^{21}$ & $2.29\times10^{20}$  \\
HD209458b & 0.3 & 1400 & 1   & $2.22\times10^{19}$ & $1.30\times10^{17}$ & $9.18\times10^{14}$  \\
HD209458b & 0.3 & 1400 & 10 & $7.01\times10^{19}$ & $4.10\times10^{17}$ & $2.89\times10^{15}$  \\
HD209458b & 0.3 & 1700 & 1   & $6.97\times10^{20}$ & $7.67\times10^{18}$ & $8.02\times10^{17}$  \\
HD209458b & 0.3 & 1700 & 10 & $2.21\times10^{21}$ & $2.43\times10^{19}$ & $1.90\times10^{18}$  \\
HD209458b & 0.3 & 2000 & 1   & $1.38\times10^{22}$ & $3.13\times10^{20}$ & $4.05\times10^{19}$  \\
HD209458b & 0.3 & 2000 & 10 & $4.52\times10^{22}$ & $1.05\times10^{21}$ & $9.42\times10^{19}$  \\
Tres-4b & 0.24 & 2000 & 1 & $6.87\times10^{22}$ & $2.57\times10^{21}$ & $1.42\times10^{20}$  \\
Tres-4b & 0.24 & 2250 & 1 & $1.44\times10^{23}$ & $3.33\times10^{21}$ & $3.68\times10^{20}$  \\
Tres-4b & 0.24 & 2500 & 1 & $4.62\times10^{23}$ & $7.87\times10^{21}$ & $1.54\times10^{21}$  \\
Tres-4b & 0.3 & 2000 & 1 & $4.80\times10^{22}$ & $9.56\times10^{20}$ & $3.16\times10^{19}$  \\
Tres-4b & 0.3 & 2250 & 1 & $1.98\times10^{23}$ & $5.92\times10^{21}$ & $6.16\times10^{20}$  \\
Tres-4b & 0.3 & 2500 & 1 & $5.13\times10^{23}$ & $8.75\times10^{21}$ & $1.55\times10^{21}$  \\
HD189733b & 0.3 & 1500 & 1 & $9.94\times10^{18}$ & $2.65\times10^{16}$ & $1.00\times10^{16}$  \\
\tableline
\end{tabular}
\end{center}
\end{table}

We approximate the electrical conductivity profile in the atmosphere with exponential functions:
\begin{eqnarray}
\sigma= \left\{\begin{array}{ll}
\sigma_{\delta}e^{\frac{r-(R-\delta)}{H_{\delta}}}\ \ \ \ &\mbox{$R-\delta<r\leqslant R$}\\
\sigma_{\gamma}e^{\frac{r-R}{H_{\gamma}}}\ \ \ \ &\mbox{$R<r\leqslant R+\gamma$}\\
\end{array}  \right.
\end{eqnarray}
where $\sigma_{\delta}$ and $\sigma_{\gamma}$ are the conductivities at $r=R-\delta$ and $r=R$ respectively, while $H_{\delta}$ and $H_{\gamma}$ are the conductivity scale-heights in the corresponding regions. We prescribe a parabolic radial dependence to the zonal flow over the thickness of the atmosphere, $\delta$, and maintain the velocity constant over the outermost thin shell, $\gamma$:
\begin{eqnarray}
\vec{v}= \left\{ \begin{array}{ll}
0\ \ \ \    &\mbox{$0<r\leqslant R-\delta$}\\
\beta  v_m\sin\theta\hat{\phi}\ \ \ \            &\mbox{$R-\delta<r\leqslant R+\gamma$}\\
\end{array}\right.
\end{eqnarray}
where
\begin{eqnarray}
\beta= \left\{\begin{array}{ll}
\left(\frac{r-(\tilde{R}-\delta)}{\delta}\right)^2\ \ \ \    &\mbox{$R-\delta<r\leqslant R$}\\
1\ \ \ \           &\mbox{$R<r\leqslant R+\gamma$}\\
\end{array}\right.
\end{eqnarray}
Assuming alignment of the dipole moment and the rotation axis, the background dipole magnetic field can be expressed as follows:
\begin{equation}
\vec{B}_{dip}=\vec{\nabla}\times k\left(\frac{\sin\theta}{r^2}\right)\hat{\phi}.
\end{equation}
With these expressions, we can decompose the angular part of $\vec{v}\times\vec{B}$ into spherical harmonics. Upon inspection, one finds that the only harmonic of interest has $\ell=2$ and $m=0$. Consequently, we write the potential as $\Phi=g(r)Y_2^0(\theta,\phi)$ and equation (7) becomes a scalar equation.

Because the outer edge of our models is set at an insulating shell, we require the radial current at $r=R+\gamma$ to be zero:
\begin{equation}
 g_{\gamma}'(R+\gamma)=\sqrt{\frac{\pi }{5}} \frac{4kv_m}{3(R+\gamma)^3}.
\end{equation}
This boundary condition is appropriate when the electrical resistance that the current will encounter radially, greatly exceeds that of a path confined to a surface i.e. $\int_{R}^{R+2\gamma}\sigma^{-1}dr\gg R\int_{0}^{\frac{\pi}{2}}\sigma^{-1}d\theta.$ This criterion is satisfied in our models.

With this boundary condition, the radial part of the solution to equation (7) in the outermost shell ($R<r\leqslant R+\gamma$) reads:
\begin{eqnarray}
g_\gamma(r)&=&\frac{e^{-\frac{R+r+\gamma}{H_\gamma}}}{90H_{\gamma}r^3\left(1-4H_\gamma+6H_\gamma^2\right)\left(12H_\gamma^2-6H_\gamma(R+\gamma)+(R+\gamma)^2\right)}\nonumber\\
&\times&(12\sqrt{5\pi}kv_me^{\frac{R+r+\gamma}{H_\gamma}}H_\gamma\left(1-4H_\gamma+6H_\gamma^2\right)(6H_\gamma^2(2R-5r+2\gamma)\nonumber\\
&-&2H_\gamma(R^2-3Rr-3r^2+2R\gamma-3r\gamma+\gamma^2)-r(r^2+(R+\gamma)^2))\nonumber\\
&-&30e^{\frac{R+\gamma}{H_\gamma}}H_\gamma^2(4H_\gamma+r)(12H_\gamma^2-6H_\gamma(R+\gamma)+(R+\gamma)^2)A_1\nonumber\\
&+&5e^{\frac{r}{H_\gamma}}\left(24H_\gamma^3-18H_\gamma^2r+6H_\gamma r^2-r^3\right)\left(12H_\gamma^2+6H_\gamma(R+\gamma)+(R+\gamma)^2\right)A_1),
\end{eqnarray}
where $A_1$ is an undetermined constant of integration. In a similar fashion, we can write down the solution to the radial part of equation (7) in the region ($R-\delta<r\leqslant R$):
\begin{eqnarray}
g_{\delta}(r)&=&\frac{1}{15r^3}(\frac{5A_2\left(-24H_{\delta}^3+18H_{\delta}^2r-6H_{\delta}r^2+r^3\right)}{6H_{\delta}^2-4H_{\delta}+1}-\frac{5A_3H_{\delta}e^{-\frac{r}{H_{\delta}}}(4H_{\delta}+r)}{6H_{\delta}^2-4H_{\delta}+1}\nonumber\\ 
&-&\frac{2\sqrt{5\pi}kv_me^{-\frac{r}{H_{\delta}}}}{\delta^2}(12H_{\delta}^2(4H_{\delta}+r)\textrm{Ei}\left(\frac{r}{H_{\delta}}\right)+e^{\frac{r}{H_{\delta}}}\nonumber\\
&\times&(-192H_{\delta}^3+96H_{\delta}^2r+2(-24H_{\delta}^3+18H_{\delta}^2r-6H_{\delta}r^2+r^3)\log (r)\nonumber\\
&-&2 H_{\delta}(12r^2+R^2-\delta^2)+r(4rR-R^2+\delta^2))),
\end{eqnarray}
where $A_2$ and $A_3$ are again undetermined constants, and \textrm{Ei} is an exponential integral: $\textrm{Ei}(x)=\int_{-\infty}^x\frac{e^t}{t}dt.$ Although equation (7) must be solved numerically in the interior, towards the center of the planet, where conductivity can be taken to be constant, it reduces to Laplace's equation. As a result we can use the polynomial eigenfunction $g_{int}(r)=A_4r^2$ in the vicinity of the origin, and $A_4$ is the last undetermined constant. The four constants of integration are determined by continuity.

\end{appendix}

\textbf{Acknowledgments} We thank G. Laughlin, D. Charbonneau, A. Showman, J. Liu, H. Knutson, A. Wolf and M. Line for useful discussions.

\end{document}